\pdfoutput=1
\documentclass[manuscript]{rmaa}
\usepackage {graphicx}
\usepackage{natbib}
\def\aap{A\&A\,  }
\def\apj{ApJ\,  }
\def\apjl{ApJ\,  }
\def\apjs{ApJS  }

\def\mnras{MNRAS\,  }

\title
{
A near infrared test  for two recent
luminosity  functions for galaxies
}
\author{L. Zaninetti\altaffilmark{1}}

\altaffiltext{1}{Dipartimento di Fisica, Via Pietro Giuria 1,
10125 Torino,  Italy (zaninetti@ph.unito.it.)}
\shortauthor{Zaninetti}
\shorttitle {Infrared luminosity function}
\ReceivedDate{...............}
\AcceptedDate{...}
\SetYear{2012}
\resumen{Este documento describe ..
 }

\abstract
{
Two recent  luminosity function (LF) for galaxies 
are reviewed and the parameters which characterize 
the near infrared are fixed.
A first LF is a 
modified Schechter LF
with four parameters.
The second LF is derived from
the generalized gamma and has  four parameters.
The formulas which give the number of galaxies 
as function of the redshift are reviewed 
and a special attention  is given to 
the position of the photometric maximum which is
expressed as function of a critical parameter or
the flux of radiation or the apparent magnitude.
A simulation of the 2MASS Redshift Survey
is given in the framework of the non Poissonian 
Voronoi Tessellation.
}
\addkeyword{Galaxies: luminosity function}
\addkeyword{Statistical distributions}
\addkeyword{Galaxies }
\addkeyword{Clusters of galaxies}

\begin{document}
\maketitle

\section{Introduction}
The release  of the 2MASS Redshift Survey (2MRS),
with it's 
44599  galaxies 
having  $K_s < 11.75$ 
allows to make tests on the radial number of galaxies 
because we have a small  zone-of-avoidance
,
see Figure 1 in  \citet{Huchra2012}.  
The number of galaxies as function of the redshift , $z$, 
is strictly related to the chosen 
luminosity function for galaxies (LF).
The most used LF is the Schechter function ,
introduced by 
\citet{schechter}
,
but also two recent  LFs, 
the generalized gamma with four parameters, see 
\citet{Zaninetti2010f}, 
and the modified Schechter LF , see 
\citet{Alcaniz2004}, can model the LF for galaxies.
We now outline  some topic issues in which the LF 
plays a relevant  role :
determination  of the parameters at different $z$ ,
 see  \citet{Goto2011},
evaluation of the parameters taking account of 
the star formation
(SF) and presence of active nuclei, 
see \citet{Wu2011},
determination of the normalization in the Near Infrared 
(NIR) as function of $z$,
see  \citet{Keenan2012}.

In this 
paper Section \ref{seclum}
first reviews  two recent LFs for galaxies 
and then derives the free parameters in the near 
infrared band. 
Once the basic parameters of the recent LFs are 
derived we make a comparison between observed radial 
distribution in the number of galaxies 
and  theoretical predictions, see Section \ref{secradial}.
A simulation of the near infrared 
all sky survey
is  reported
in Section \ref{simulation}.

\section{The luminosity functions}

\label{seclum}

\label{sec_lfs}
This Section  reviews
the  standard luminosity function (LF)
for galaxies,
and two  recent
LF  for galaxies.
A first test is done on Table 2 of \citet{Cole2001}
where the combined data of
 the Two Micron All Sky Survey (2MASS) Extended Source Catalog 
and the 2dF Galaxy Redshift Survey
 allowed to build a luminosity function in the
$K_s$  band (2MASS Kron magnitudes).
The main statistical test is done through
the $\chi^2$,
\begin{equation}
\chi^2 = \sum_{i=1}^n \frac { (T_i - O_i)^2} {T_i},
\label{chisquare}
\end {equation}
where $n  $   is the number of bins,
      $T_i$   is the theoretical value,
and   $O_i$   is the experimental value represented
by the frequencies.
A reduced  merit function $\chi_{red}^2$
is  evaluated  by
\begin{equation}
\chi_{red}^2 = \chi^2/NF
\quad,
\label{chisquarereduced}
\end{equation}
where $NF=n-k$ is the number of degrees  of freedom,
$n$     is the number of bins,
and $k$ is the number of parameters.

\subsection{The  Schechter function }

The  Schechter function, 
introduced by
\citet{schechter},
provides a useful fit  for the
LF  of galaxies
\begin{equation}
\Phi (L) dL  = (\frac {\Phi^*}{L^*}) (\frac {L}{L^*})^{\alpha}
\exp \bigl ( {-  \frac {L}{L^*}} \bigr ) dL \quad  ,
\label{equation_schechter}
\end {equation}
here $\alpha$ sets the slope for low values
of $L$ , $L^*$ is the
characteristic luminosity and $\Phi^*$ is the normalization.
The equivalent distribution in absolute magnitude is
\begin{equation}
\Phi (M)dM=0.921 \Phi^* 10^{0.4(\alpha +1 ) (M^*-M)}
\exp \bigl ({- 10^{0.4(M^*-M)}} \bigr)  dM \, ,
\label{lfstandard}
\end {equation}
where $M^*$ is the characteristic magnitude as derived from the
data.
The scaling with  $h$ is  $M^* - 5\log_{10}h$ and
$\Phi^* ~h^3~[Mpc^{-3}]$.

\subsection{A modified  Schechter function }

In order  to  improve the flexibility
at the bright  end
\citet{Alcaniz2004}  introduced
a new parameter
$\eta$ in the Schechter LF
\begin{equation}
\Phi (L) dL  =
\frac
{
{\it \Phi^*}\, \left( {\frac {L}{{\it L^*}}} \right) ^{\alpha}
 \left( 1-{\frac { \left( \eta-1 \right) L}{{\it L^*}}} \right)
^{\frac{1}{ \eta-1 }}
}
{
{\it L^*}
}
dL
\quad  .
\end{equation}
This new LF in the case  of
$\eta < 1$ is defined in the range
$0 < L < L_{max} $
where
$L_{max}=\frac{L^*}{\eta-1} $
and therefore has a natural upper boundary  which is
not infinity.
In the limit $\lim_{\eta \to 1} \Phi (L)$ the
Schechter LF is obtained.
In the case of $\eta > 1$ the average value is
\begin{equation}
{ \langle L \rangle }
=
\frac
{{\it \Phi^*}{\it L^*}\,
\Gamma  \left( 3+\alpha \right) \Gamma  \left( 1+ \left( \eta-1
 \right) ^{-1} \right)  \left( \eta-1 \right) ^{-\alpha-2
}
}
{
\Gamma  \left( 3+\alpha+ \left( \eta-1 \right) ^{-1} \right)  \left(
\alpha+2 \right)
}
\quad ,
\end{equation}
and in the case of $\eta <1 $ the average value is
\begin{equation}
{ \langle L \rangle }
=
\frac
{
{\it \Phi^*}{\it L^*}\,
\left( -\eta+1 \right) ^{-2-\alpha}\,\Gamma
 \left( -{\frac {-1+2\,\eta+\alpha\,\eta-\alpha}{\eta-1}} \right)
\Gamma  \left( 2+\alpha \right)
}
{
\Gamma  \left( - \left( \eta-1 \right) ^{-1} \right)
}
\quad .
\end{equation}

\begin{figure}
 \centering
\includegraphics[width=10cm]{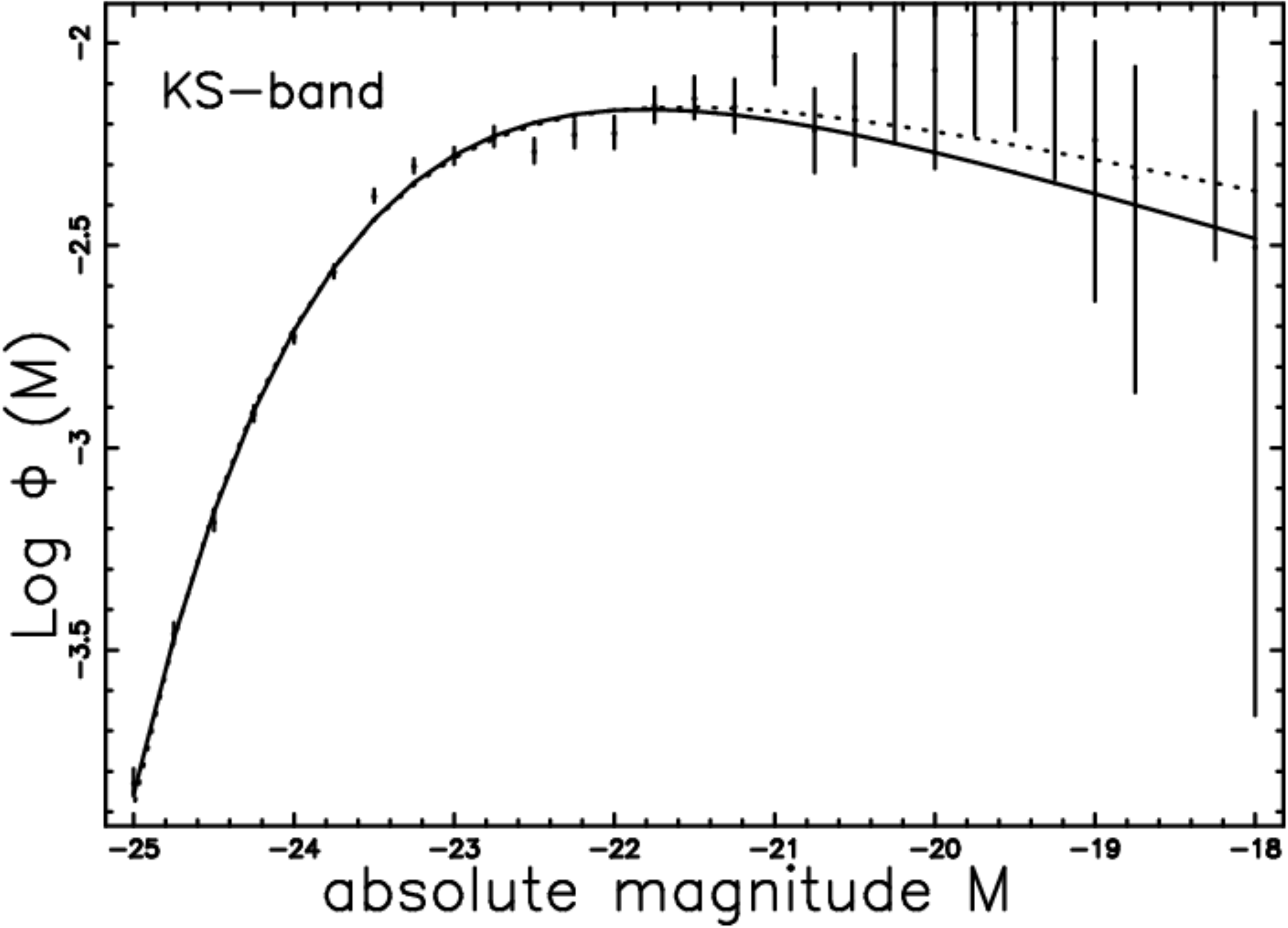}
\caption 
{
The luminosity function data of the 
$K_S$ band are represented with error bars.
The continuous line fit represents the modified Schechter LF
(eqn. \ref{lfbrazilian})
and the dotted
line represents the Schechter LF (eqn.\ref{lfstandard}).
 }
 \label{due_brazilian}
 \end{figure}

The distribution  in magnitude is :

\begin{equation}
\Phi(M)dM=
 0.921
 \Phi^*
10^{0.4(M^*-M) (\alpha+1)}
(1- (\eta -1) 10^{0.4(M^*-M)} )
^{\frac{1}{ \eta-1 }}
dM
\quad  .
\label{lfbrazilian}
\end{equation}
Table \ref{chi2value} reports  the
parameters of the Schechter and the  modified Schechter LF
applied to the $K_s$ band.
\begin{table}[ht!]
\caption { Numerical values values and  $\chi^2_{red}$ of the three LFs
applied to $K_S$ band
(2MASS Kron magnitudes) with
data as extracted from Table 2 in Cole et al. 2001
when $M_{\sun}$=3.39 \,. }
\label{chi2value}
\begin{center}
\begin{tabular}{|c|c|c|}
\hline
LF        &   parameters    & $\chi^2_{red}$ \\
\hline
Schechter, Coles~2001  &
$M^*$= -23.77, $\alpha$=-1.14,
 &
 ~ \\
 ~ &  $\Phi^* = 0.0114 /Mpc^3$ &~ \\
 \hline
 Schechter, our code  &
$M^*$= -23.289, $\alpha$=-0.794,
 &
 1.38 \\
 ~ &  $\Phi^* = 0.0128 /Mpc^3$ & ~  \\
 \hline
 Modified Schechter &
$M^*$= -23.119, $\alpha$= -0.707, &
 1.36 \\
 ~ & $\Phi^* =   0.0142 /Mpc^3$
 , $\eta=0.956$   & ~ \\
 \hline
 Generalized gamma  &
$M^*$= -23.14 , k=0.946,c=0.28 &
  1.42 \\
  ~ & $\Psi^* =    0.0481/Mpc^3$ & ~ \\
 \hline
\end{tabular}
\end{center}
\end{table}
The Schechter LF, the  
modified Schechter LF 
as well the observed data in the $K_S$ band  are
reported in
Figure~\ref{due_brazilian}.

\subsection{The generalized gamma distribution with four parameters}

A  four parameter generalized gamma LF
has  been derived
in \citet{Zaninetti2010f}
\begin{equation}
\Psi(L;L^*,c,k,\Psi^*)=
\Psi^*\frac
{
{\it k}\, \left( {\frac {L}{L^*}} \right) ^{c{\it k}-1}{{\rm e}^{-
 \left( {\frac {L}{L^*}} \right) ^{{\it k}}}}
}
{
L^*\Gamma \left( c \right)
}
\quad .
\label{lf4}
\end{equation}
This function contains the four parameters $c$,
$k$, $L^*$
and $\Psi^*$ and
the range of existence is $ 0 \leq L < \infty $.
The averaged luminosity is
\begin{equation}
{ \langle L \rangle }
=
\frac
{
{\it L^*}\,\Gamma \left( {\frac {1+ck}{k}} \right)
}
{
\Gamma \left( c \right)
}
\quad .
\label{lmedio4}
\end{equation}
The magnitude version of this  LF is
\begin{equation}
\Psi (M) dM =
\frac
{
{ \Psi^*}0.4\ln \left( 10 \right)k{10}^{- 0.4ck \left( M- { M^*}
 \right) }{{\rm e}^{- {10}^{- 0.4 \left( M- {
M^*} \right) k}}}
}
{
\Gamma \left( c \right)
}
\nonumber   \\
 dM \, .
\label{lfgamma}
\end {equation}

\begin{figure}
 \centering
\includegraphics[width=10cm]{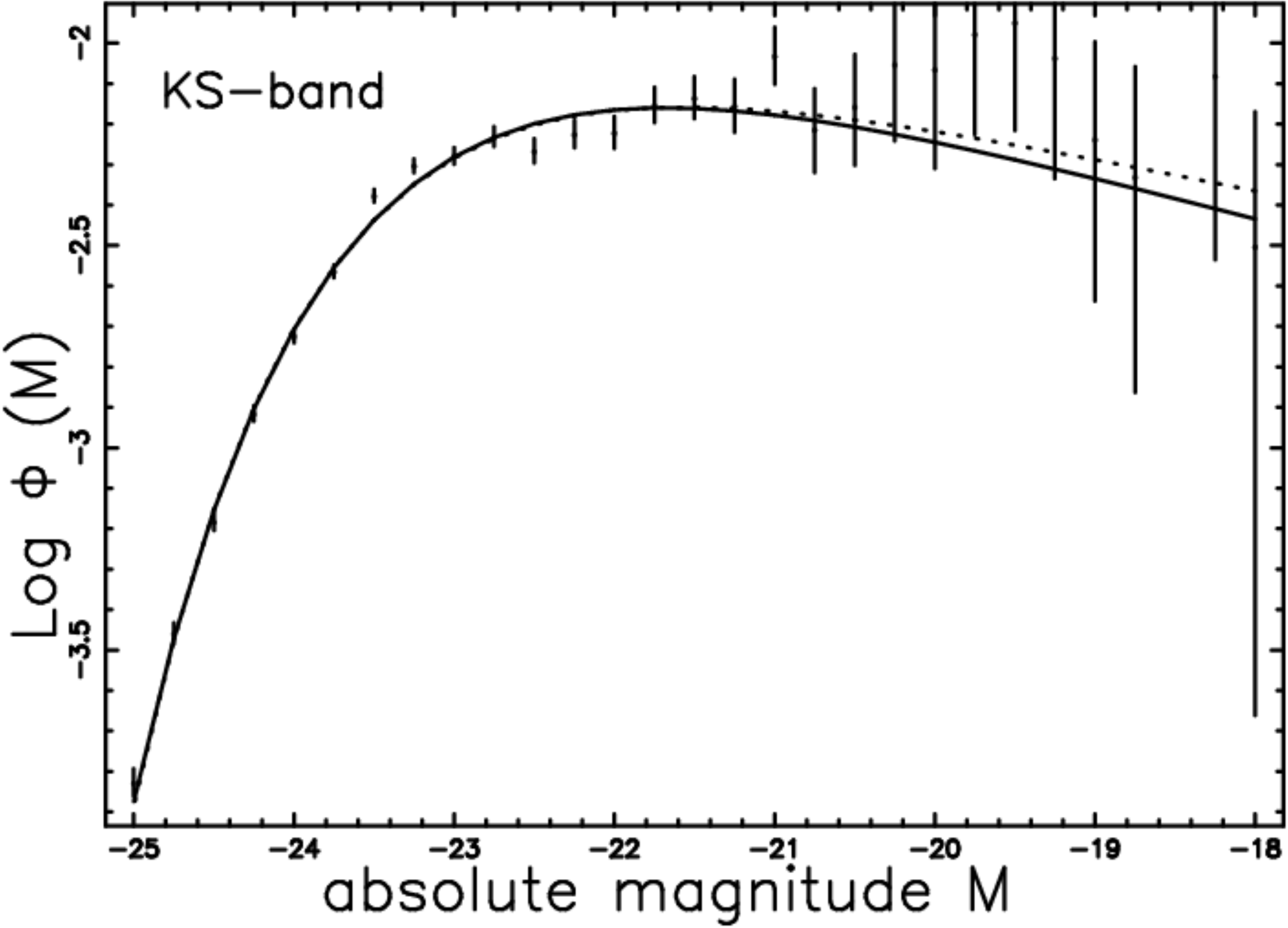}
\caption 
{
The luminosity function data of the 
$K_S$ band are represented with error bars.
The continuous line fit 
represents the 
generalized gamma distribution
 LF
(eqn. \ref{lfgamma})
and the dotted
line represents the Schechter LF (eqn.\ref{lfstandard}).
}
 \label{due_gamma}
 \end{figure}

The Schechter LF, the  
generalized gamma LF 
as well the observed data in the $K_S$ band  are
reported in
Figure~\ref{due_gamma}
and the adopted parameters in Table \ref{chi2value}. 

\section{Number of galaxies and redshift}

\label{secradial}

In this section  we processed  the 
2MASS Redshift Survey (2MRS), 
see \citet{Huchra2012}.  

\subsection{The existing formulas}

We assume that the  correlation
between expansion velocity  and distance is
\begin {equation}
V= H_0 D  = c_l \, z  
\quad  ,
\label {clz}
\end{equation}
where $H_0$ is the Hubble constant , after \citet{Hubble1929},
$H_0 = 100 h \mathrm{\ km\ s}^{-1}\mathrm{\ Mpc}^{-1}$, with $h=1$
when  $h$ is not specified,
$D$ is the distance in $Mpc$,
$c_l$ is  the  light velocity  and
$z$   
is the redshift.
Concerning   the exact value of  $H_0$ 
a recent value as obtained by
the 
 mid-infrared calibration of the 
Cepheid distance scale based on
recent observations at 3.6 $\mu m$ , see 
\citet{Freedman2012}, 
suggests 
\begin{equation}
H_0 =(74.3 \pm 2.1 ) \mathrm{\ km\ s}^{-1}\mathrm{\ Mpc}^{-1}
\quad .
\end {equation}
In an Euclidean ,non-relativistic 
and homogeneous universe 
the flux of radiation,
$ f$,  expressed in $ \frac {L_{\sun}}{Mpc^2}$ units,
where $L_{\sun}$ represents the luminosity of the sun
,  is 
\begin{equation}
f  = \frac{L}{4 \pi D^2}  
\quad ,
\end{equation}
where $D$   represents the distance of the galaxy 
expressed in $Mpc$,
and  
\begin{equation}
D=\frac{c_l z}{H_0} 
\quad  .
\end{equation}

The relationship connecting the absolute magnitude, $M$ ,
 of a
galaxy  to  its luminosity is
\begin{equation}
\frac {L}{L_{\sun}} =
10^{0.4(M_{\sun} - M)}
\quad ,
\label{mlrelation}
\end {equation}
where $M_{\sun}$ is the reference magnitude 
of the sun at the considered bandpass.

The flux   expressed in $ \frac {L_{\sun}}{Mpc^2}$ units 
as  a function of the  
apparent magnitude is
\begin{eqnarray}
f=
7.957 \times 10^8 \,{e^{ 0.921\,{\it M_{\sun} }- 0.921\,{\it
m}}}
\quad    \frac {L_{\sun}}{Mpc^2} \quad , 
\label{damaf}
\end {eqnarray}
and  the inverse relationship is 
\begin{eqnarray}
m=
M_{\sun}- 1.0857\,\ln  \left(  0.1256 \times 10^{-8} f \right) 
\quad . 
\label{dafam} 
\end {eqnarray}
The joint distribution in {\it z}  
and {\it f}  for galaxies ,
see formula~(1.104) in
 \citet{pad} 
or formula~(1.117) 
in 
\citet{Padmanabhan_III_2002} 
,
 is
\begin{equation}
\frac{dN}{d\Omega dz df} =  
4 \pi  \bigl ( \frac {c_l}{H_0} \bigr )^5    z^4 \Phi (\frac{z^2}{z_{crit}^2})
\label{nfunctionzschechter}  
\quad ,
\end {equation}
where $d\Omega$, $dz$ and  $ df $ represent 
the differential of
the solid angle, 
the redshift and the flux respectively
and     $\Phi$ is the Schechter LF.
The critical value of $z$,   $z_{crit}$, is 
\begin{equation}
 z_{crit}^2 = \frac {H_0^2  L^* } {4 \pi f c_l^2}
\quad .
\end{equation} 
The number of galaxies in $z$  and $f$ as given by 
formula~(\ref{nfunctionzschechter})  
has a maximum  at  $z=z_{pos-max}$ ,
where 
\begin{equation}
 z_{pos-max} = z_{crit}  \sqrt {\alpha +2 }
\quad ,
\end{equation} 
which  can be re-expressed   as
\begin{equation}
 z_{pos-max}(f) =
\frac
{
\sqrt {2+\alpha}\sqrt {{10}^{ 0.4\,{\it M_{\sun}}- 0.4\,{\it M^*}}}{
\it H_0}
}
{
2\,\sqrt {\pi }\sqrt {f}{\it c_l}
}
\quad  ,
\label{zmax_sch}
\end{equation}
or  replacing  the flux $f$ with the apparent  
magnitude $m$ 
\begin{equation}
 z_{pos-max}(m) =
\frac
{
1.772\,10^{-5}
\,\sqrt {2+\alpha}\sqrt {{10}^{ 0.4\,M_{{{\it \sun}}}-
 0.4\,{\it M^*}}}H_{{0}}
}
{
\sqrt {\pi }\sqrt {{{\rm e}^{ 0.921\,M_{{{\it \sun}}}-
 0.921\,m}}}{\it c\_l}
}
\quad  .
\label{zmax_schmag}
\end{equation}

The number of galaxies, $N_S(z,f_{min},f_{max})$  
comprised between a minimum value of flux,
 $f_{min}$,  and  maximum value of flux $f_{max}$,
can be computed  through  the following integral 
\begin{equation}
N_S (z) = \int_{f_{min}} ^{f_{max}}
4 \pi  \bigl ( \frac {c_l}{H_0} \bigr )^5    z^4 \Phi (\frac{z^2}{z_{crit}^2})
df
\quad .
\label{integrale} 
\end {equation}

\subsection{Formulas for the modified Schechter LF}

\begin{figure}
\begin{center}
\includegraphics[width=6cm]{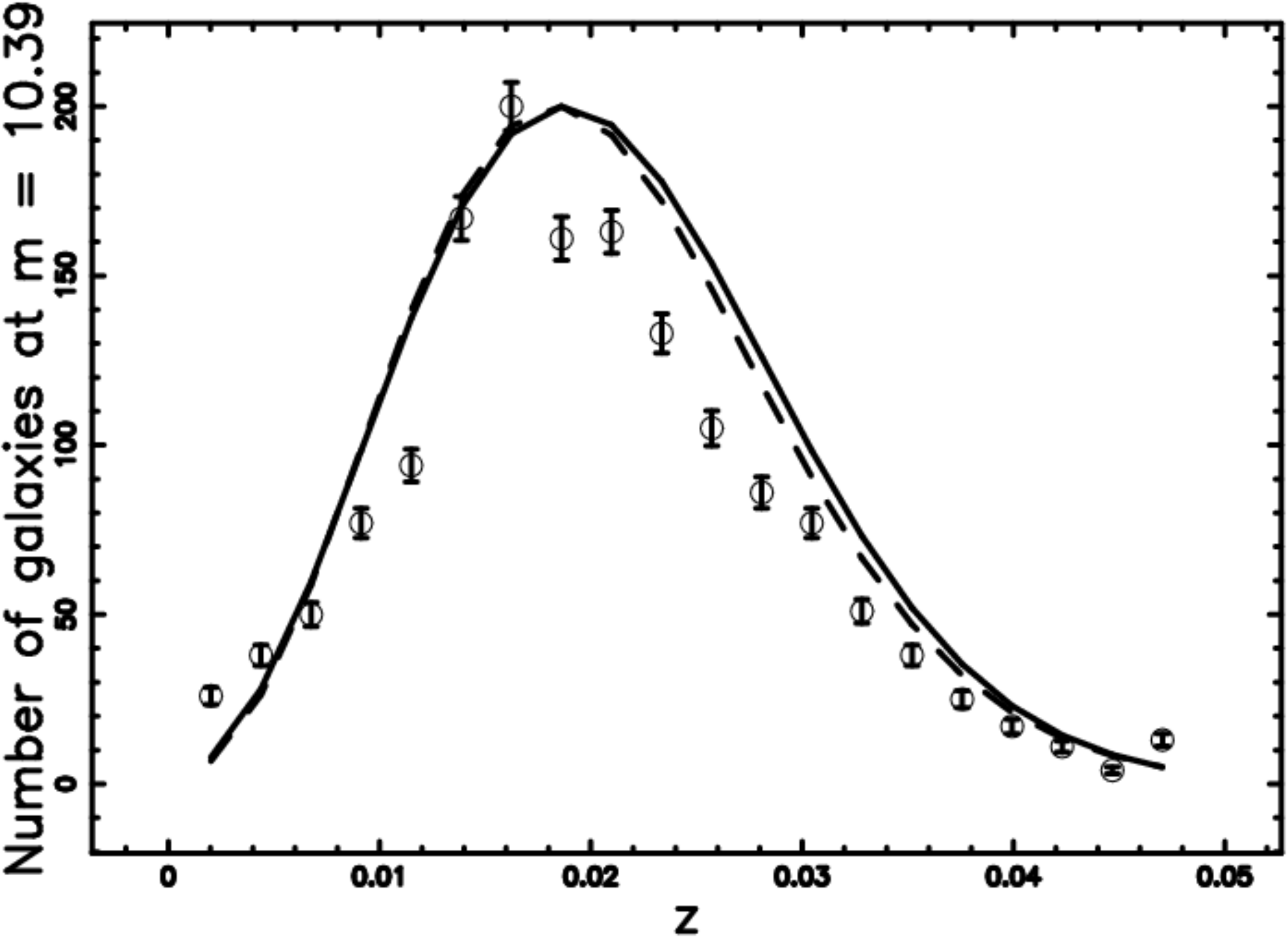}
\end {center}
\caption{
The galaxies  of the 2MRS with
$ 10.28    \leq  m   \leq 10.44  $  or
$ 1202409  \frac {L_{\sun}}{Mpc^2} \leq
f \leq  1384350 \frac {L_{\sun}}{Mpc^2}$
are  organized in frequencies versus
heliocentric  redshift,
(empty circles);
the error bar is given by the square root of the frequency.
The maximum frequency of observed galaxies is
at  $z=0.018$.
The full line is the theoretical curve  
generated by  
$\frac{dN}{d\Omega dz df}(z)$
as given by the application of the Schecter LF
which  is eqn.(\ref{nfunctionzschechter})
and the dashed line  
represents the application of the Schechter LF
which  is eqn.(\ref{nfunctionzmodifiedschechter}).
The parameters are  the same of Table \ref{chi2value},
$\chi^2= 154$  for the Schechter LF    and 
$\chi^2= 124$  for the modified Schechter LF.
}
          \label{maximum_flux_brazilian}%
    \end{figure}

\begin{figure}
\begin{center}
\includegraphics[width=6cm]{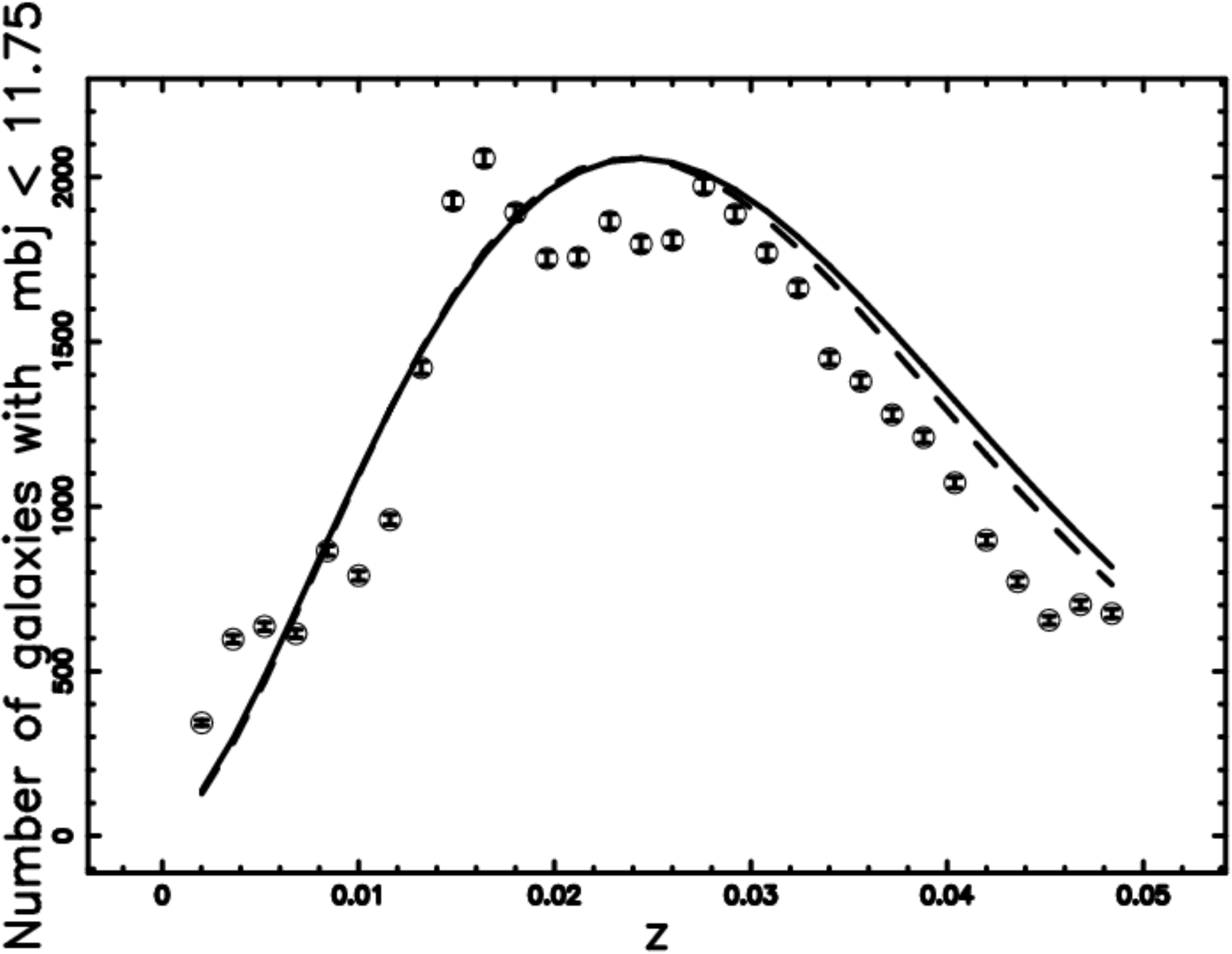}
\end {center}
\caption{
The same  as  Figure  \ref{maximum_flux_brazilian}
but now all the galaxies are considered,
$\chi^2= 1618$  for the Schechter LF    and 
$\chi^2= 1352$  for the modified Schechter LF.
}
          \label{maximum_flux_tutte_brazilian}%
    \end{figure}

The joint distribution in 
$z$
and $f$  
for galaxies  when the 
modified Schechter LF is adopted is 
\begin{equation}
\frac{dN}{d\Omega dz df} = 
\frac
{
4\,{z}^{4}{c_{{l}}}^{5}{\it \Phi^*}\, \left( {\frac {{z}^{2}}{{z_{{{
\it crit}}}}^{2}}} \right) ^{\alpha} \left( -{\frac {-{z_{{{\it crit}}
}}^{2}+{z}^{2}\eta-{z}^{2}}{{z_{{{\it crit}}}}^{2}}} \right) ^{
 \left( \eta-1 \right) ^{-1}}\pi 
}
{
{H_{{0}}}^{5}{\it L^*}
}
\label{nfunctionzmodifiedschechter}  
\quad .
\end {equation}
Figure \ref{maximum_flux_brazilian}
reports the number of  observed  galaxies
of the 2MRS  catalog for  a given   
apparent magnitude  and
two theoretical curves.
The number   of  all the galaxies
as  function of $z$  can be computed
through an integral, see eqn. \ref{integrale},
and  is visible  in Figure \ref{maximum_flux_tutte_brazilian}.

The maximum in the number of galaxies 
is at 
\begin{equation}
 z_{pos-max}(z_{crit}) = 
\frac
{
\sqrt {2+\alpha}{\it z_{crit}}
}
{
\sqrt {2\,\eta-1+\alpha\,\eta-\alpha}
}
\quad ,
\label{zmax_mod1}
\end{equation} 
or 
\begin{equation}
 z_{pos-max}(f) =
\frac
{
\sqrt {2+\alpha}\sqrt {{10}^{ 0.4\,{\it M_{\sun}}- 0.4\,{\it M^*}}}H_
{{0}}
}
{
2\,\sqrt {2\,\eta-1+\alpha\,\eta-\alpha}\sqrt {\pi }\sqrt {f}{\it 
c_l}
}
\quad  ,
\label{zmax_mod2}
\end{equation}

or 
\begin{equation}
 z_{pos-max}(m) =
\frac
{
1.772\,10^{-5}\,\sqrt {2+\alpha}\sqrt {{10}^{ 0.4\,{\it M_{\sun} }
-
 0.4\,{\it M^*}}}{\it H_0}
}
{
\sqrt {2\,\eta-1+\alpha\,\eta-\alpha}\sqrt {\pi }
\sqrt {{{\rm e}^{
 0.921\,{\it M_{\sun}}- 0.921 \,{\it m }}}}{\it 
c_l}
}
\quad  .
\label{zmax_mod3}
\end{equation}

\subsection{Formulas for the generalized gamma LF}

The joint distribution in 
{\it z}  
and {\it f}  
for galaxies  when the 
generalized gamma LF is adopted is 
\begin{equation}
\frac{dN}{d\Omega dz df} = 
\frac
{
4\,{z}^{4}{{\it c_l}}^{5}{\it k}\, \left( {\frac {{z}^{2}}{{{\it 
z_{crit}}}^{2}}} \right) ^{{\it c}\,{\it k}-1}{{\rm e}^{- \left( {
\frac {{z}^{2}}{{{\it z_{crit}}}^{2}}} \right) ^{{\it k}}}}{\it \Psi^*}
\,\pi 
}
{
{H_{{0}}}^{5}{\it L^*}\,\Gamma  \left( {\it c} \right) 
}
\label{nfunctionzgamma4}  
\quad .
\end {equation}

\begin{figure}
\begin{center}
\includegraphics[width=6cm]{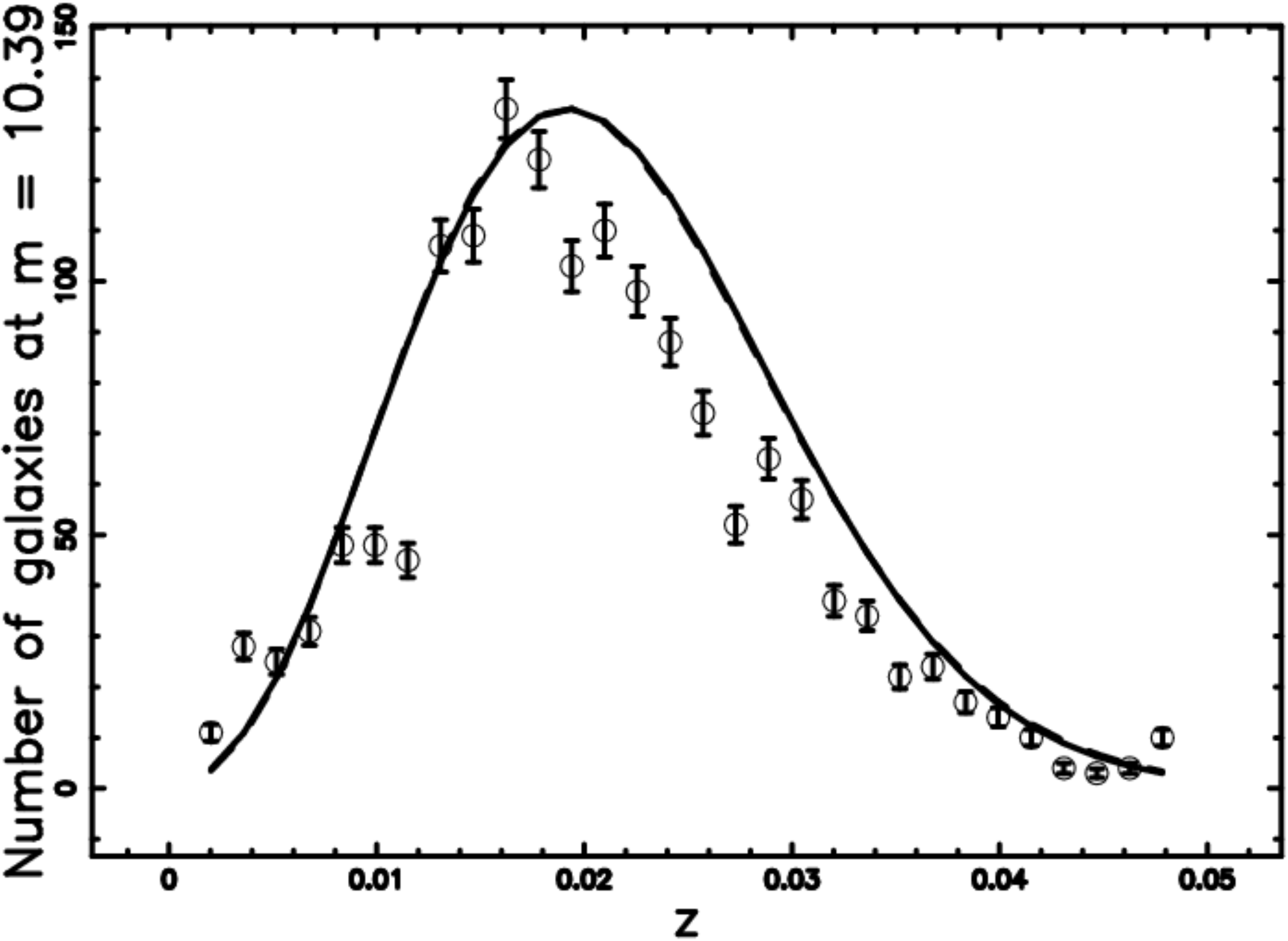}
\end {center}
\caption{
The same as Figure \ref{maximum_flux_brazilian},
the dashed line is the theoretical curve  
generated by  
$\frac{dN}{d\Omega dz df}(z)$
as given by the application of generalized gamma LF
which  is eqn. (\ref{nfunctionzgamma4}).
The parameters are  the same of Table \ref{chi2value},
$\chi^2= 198$  for the Schechter LF    and 
$\chi^2= 201$  for the generalized gamma LF.
}
          \label{maximum_flux_gamma4}%
    \end{figure}

Figure \ref{maximum_flux_gamma4}
reports the number of  observed  galaxies
of the 2MRS  catalog for  a given   
apparent magnitude  and
two theoretical curves.

\begin{figure}
\begin{center}
\includegraphics[width=6cm]{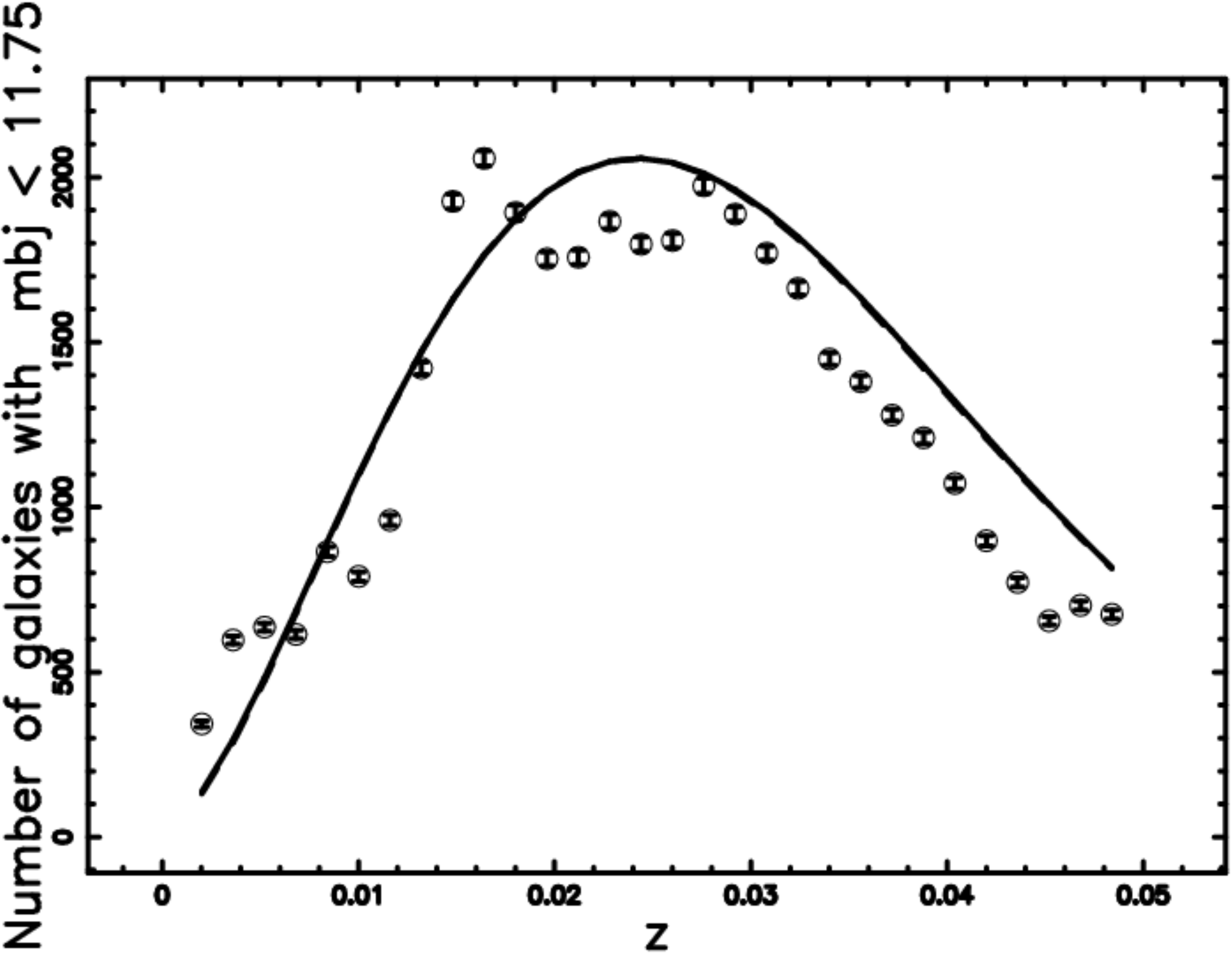}
\end {center}
\caption
{
The same  as  Figure  \ref{maximum_flux_gamma4}
but now all the galaxies are considered,
$\chi^2= 1617$  for the Schechter LF    and 
$\chi^2= 1590$  for the generalized gamma LF.
}
          \label{maximum_flux_tutte_gamma4}%
    \end{figure}

Figure \ref{maximum_flux_tutte_gamma4}
reports all the  observed  galaxies
of the 2MRS.   

The maximum in the number of galaxies is at 
\begin{equation}
 z_{pos-max}(z_{crit}) = 
{\it z_{crit}}\, \left( 1+{\it c}\,{\it k} \right) ^{1/2\,{{\it k}}^{-
1}}{{\it k}}^{-1/2\,{{\it k}}^{-1}}
\quad ,
\label{zmax_gamma1}
\end{equation} 
or 
\begin{equation}
 z_{pos-max}(f) =
\frac
{
\sqrt {{10}^{ 0.4\,{\it M_{\sun}}- 0.4\,{\it M^*}}}H_{{0}} 
\left( 1+{
\it c}\,{\it k} \right) ^{1/2\,{{\it k}}^{-1}}{{\it k}}^{-1/2\,{{
\it k}}^{-1}}
}
{
\,\sqrt {\pi }\sqrt {f}{\it c_l}
}
\quad  ,
\label{zmax_gamma2}
\end{equation}
or 
\begin{equation}
 z_{pos-max}(m) =
\frac
{
1.772\,10^{-5}\,\,\sqrt {{10}^{ 0.4\,{\it M_{\sun}}- 
0.4\,{\it M^*}}
}H_{{0}} \left( 1+{\it c}\,{\it k} \right) ^{1/2\,{{\it k}}^{-1}}{{
\it k}}^{-1/2\,{{\it k}}^{-1}}
}
{
\sqrt {\pi }\sqrt {{{\rm e}^{ 0.921\,{\it M_{\sun}}- 0.921
\,{\it m}}}}{\it c_l}
}
\quad  .
\label{zmax_gamma3}
\end{equation}

\section{The simulation}
\label{simulation}
\begin{figure}
\begin{center}
\includegraphics[width=6cm]{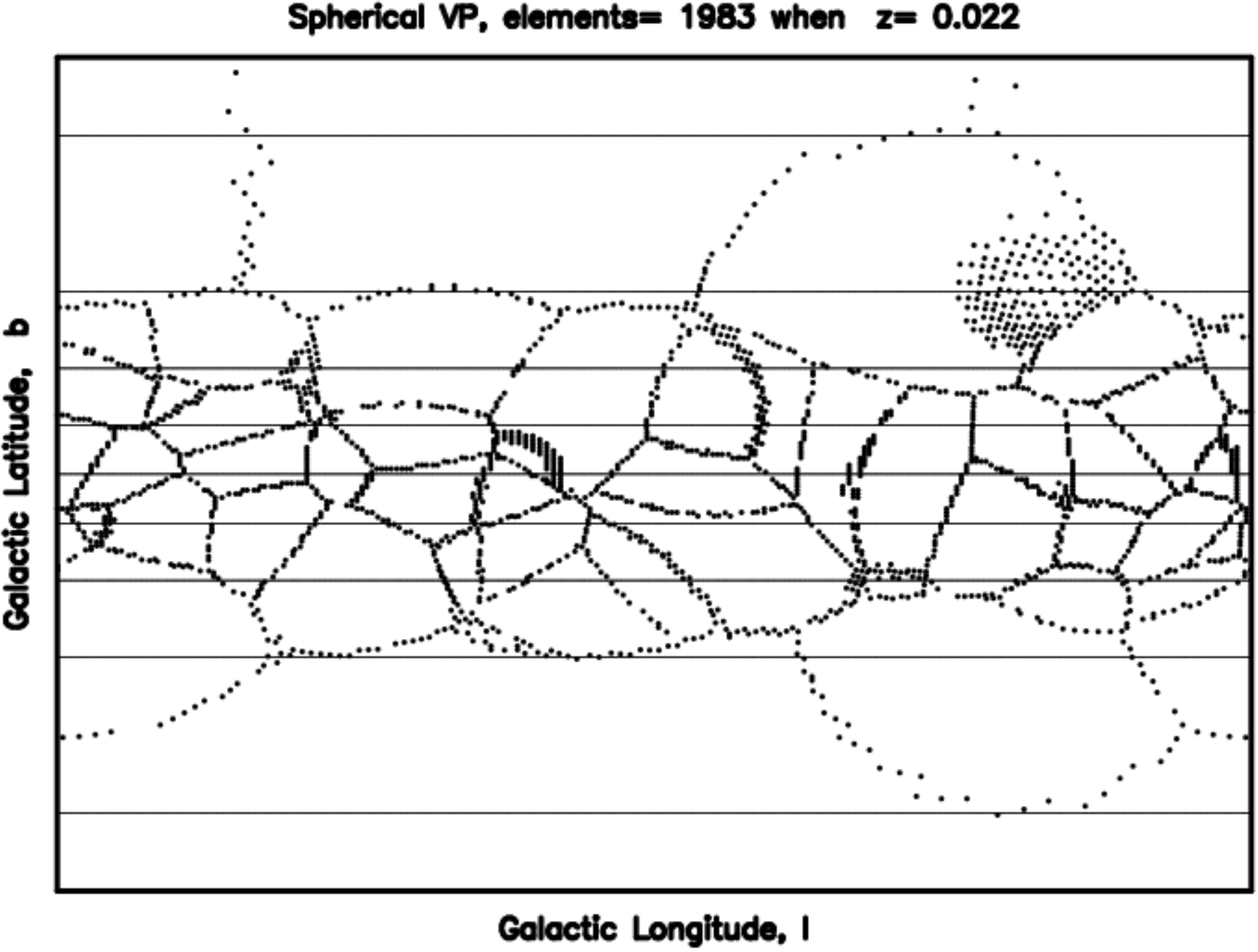}
\end {center}
\caption
{
The Voronoi diagram $V_s(2,3)$
in Mercator  projection
when $z \approx   0.022$.
Nine lines of constant latitude at latitudes -80, -60, -40,-20,
0, 20, 40, 60, 80 degrees are drawn. 
Each line is made up of 360 straight-line segments.
}
          \label{sphere_voronoi}%
    \end{figure}

\begin{figure}
\begin{center}
\includegraphics[width=6cm]{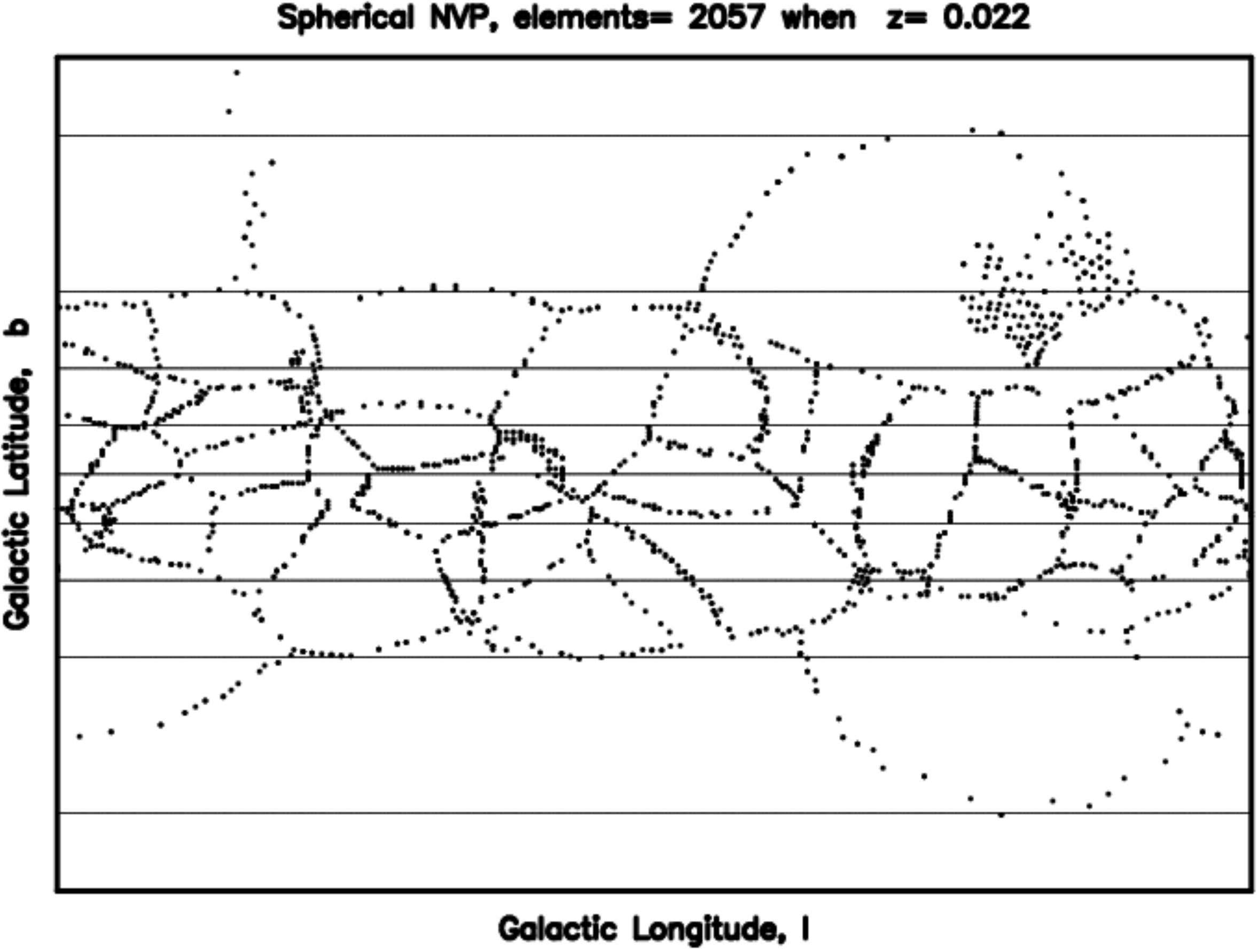}
\end {center}
\caption
{
The Voronoi diagram $V_s(2,3)$
in  Mercator  projection
when 2057 galaxies are 
extracted  from the network
of  Figure \ref{sphere_voronoi}.
}
          \label{sphere_voronoi_galaxies}%
    \end{figure}

We now  simulate the  2MRS catalog adopting the framework 
of the Voronoi Tessellation adopting two requirements.
The {\it first } requirement is that 
the average radius of the voids is 
$<R>=18.23 h^{-1}$\ Mpc,
which is  the effective radius in  SDSS DR7, see Table 6  in 
in \citet{Zaninetti2012e}.
The {\it second} requirement 
is connected to a previous analysis
which shows that   
the effective  radius  of the cosmic
voids as  deduced from
the  catalog SDSS R7  is  represented  by a Kiang function
with $c \approx 2$.
This  mean that  we are considering 
non Poissonian Voronoi
Tessellation (NPVT).
We briefly recall  that 
the Poissonian Voronoi
Tessellation (PVT) 
is characterized
by a distribution of normalized volumes
modeled by a Kiang function
with $c \approx 5$, see  \citet{Zaninetti2012e}.
The cross sectional area of a  NPVT can
also be visualized through
a spherical cut characterized by a constant value
of $z$
see Figure~\ref{sphere_voronoi};
this intersection is called $V_s(2,3)$  where the
index $s$ stands for sphere.
On this  spherical network the galaxies 
are chosen according to formula 
(\ref{integrale}) 
which represents the number of galaxies, 
$N_S(z,f_{min},f_{max})$,  
comprised between a minimum value of flux 
and  maximum value of flux
when  the Schechter LF is considered.
We have now  a series of simulated spherical 
cuts which can be compared
with the spherical  cuts of 2MRS.
Figure~\ref{sphere_voronoi_galaxies} reports
the simulated spherical slice of galaxies at the
photometric maximum. 

\begin{figure}
\begin{center}
\includegraphics[width=6cm]{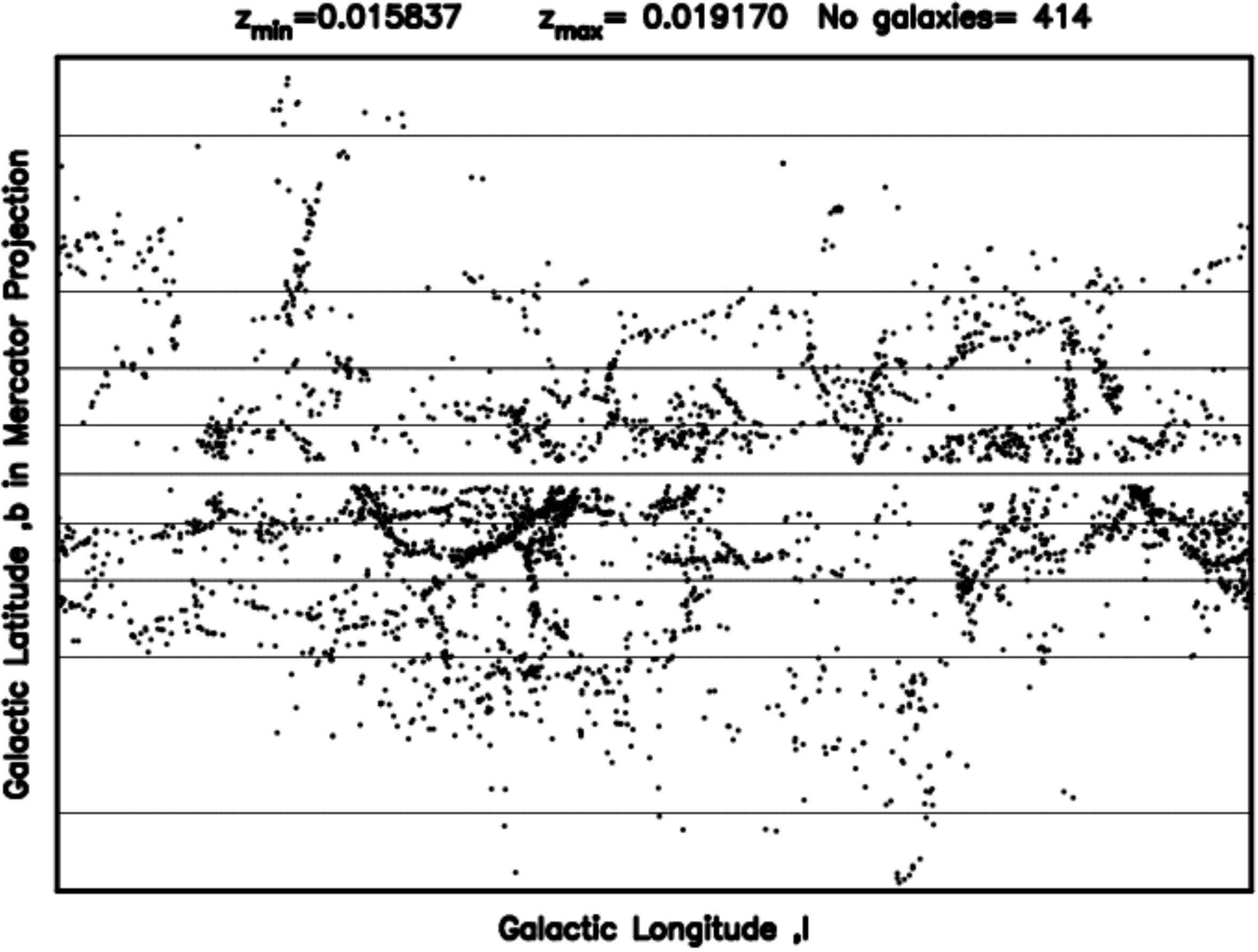}
\end {center}
\caption
{
Mercator projection in galactic coordinates
of a spherical cut of the 2MRS   data 
at $z \approx  0.017$.
}
          \label{sphere_2mrs}%
    \end{figure}

A comparison can be done with the observed spherical cut
of the 2MRS catalog having redshift ,$z$,
corresponding to the observed photometric
maximum, see Figure \ref{sphere_2mrs}.

\section{Conclusions}

The most used LF in the infrared band is  
the Schechter LF, see  eqn. (\ref{lfstandard}).
We have here analyzed  two other LFs : the generalized gamma, 
see  eqn. (\ref{lfgamma}) 
and the modified 
Schechter LF, eqn. (\ref{lfbrazilian}).
They perform as well as the Schechter LF and the 
test in the $K_s$ band  assigns to the modified
Schechter LF the lowest value of the  
$\chi^2_{red}$, see Table \ref{chi2value}.
The radial distribution in the number of galaxies 
of 2MRS
can be another test 
and Figure \ref{maximum_flux_brazilian} and 
Figure \ref{maximum_flux_gamma4} 
report the standard theoretical curve as well 
the two new theoretical predictions. 
In this case the smaller $\chi^2$ is  given by the
theoretical curve which involves the modified 
Schechter LF.
A particular attention has been given to the 
position of the maximum
in the number of galaxies that is  
here expressed as function of the
theoretical parameter 
$z_{crit}$ or the two observable parameters $f$ and $m$.
A simulation in Mercator  projection 
of the spatial distribution 
of galaxies having redshift  
corresponding to the photometric maximum is presented,
see Figure \ref{sphere_voronoi_galaxies} .


\end{document}